\documentclass[10pt]{article}
\usepackage{amsmath}
\usepackage{amssymb}
\usepackage{graphicx}
\usepackage{cite}
\usepackage[labelfont=bf,labelsep=period,justification=raggedright]{caption}
\makeatletter
\renewcommand{\@biblabel}[1]{\quad#1.}
\makeatother
\date{}
\begin{document}
\begin{flushleft}
{\Large
\textbf{In vivo facilitated diffusion model}
}
\\
Maximilian Bauer$^{1,2}$,  
Ralf Metzler$^{1,3,\ast}$
\\
\bf{1} Institute of Physics and Astronomy, Potsdam University, Potsdam-Golm, Germany 
\\
\bf{2} Physics Department, Technical University of Munich, Garching, Germany
\\
\bf{3} Physics Department, Tampere University of Technology, Tampere, Finland
\\
$\ast$ E-mail: rmetzler@uni-potsdam.de
\end{flushleft}
\section*{Abstract}
Under dilute in vitro conditions transcription factors rapidly locate their
target sequence on DNA by using the facilitated diffusion mechanism. However,
whether this strategy of alternating between three-dimensional bulk diffusion
and one-dimensional sliding along the DNA contour is still beneficial in the
crowded interior of cells is highly disputed. Here we use a simple model for
the bacterial genome inside the cell and present a semi-analytical model for the in
vivo target search of transcription factors within the facilitated diffusion
framework. Without having to resort to extensive simulations we determine the
mean search time of a lac repressor in a living \emph{E. coli} cell by
including parameters deduced from experimental measurements. The results
agree very well with experimental findings, and thus the facilitated diffusion
picture emerges as a quantitative approach to gene regulation in living
bacteria cells. Furthermore we see
that the search time is not very sensitive to the parameters characterizing
the DNA configuration and that the cell seems to operate very close to
optimal conditions for target localization. Local searches as implied by
the colocalization mechanism are only found to mildly accelerate the mean
search time within our model.

\section*{Introduction}
Transcription factors (\emph{TFs}) are able to locate and bind their target
sequence on DNA at surprisingly high rates. This became clear when
in 1970 it was measured that \emph{in vitro} the lac repressor associates
with the operator at a rate of $k_a=7\times 10^9 M^{-1}s^{-1}$
\cite{Riggs1970JMB}. This is about two orders of magnitude faster than a rate
calculated with the well-known Smoluchowski formula for three-dimensional
diffusion control \cite{Smoluchowski1916PZ}.
The results obtained in the \emph{in vitro} experiments by Riggs et~al.
and by Winter et~al. were successfully explained with the by now classical facilitated
diffusion model, introduced by Berg, von Hippel and co-workers
\cite{Berg1981BC,Winter1981BC}: the TF alternates between three-dimensional
diffusion through the bulk solution and sliding along the DNA contour which
can be considered as one-dimensional diffusion. While a large majority of subsequent
reformulations of this target search problem are based on this facilitated
diffusion model \cite{Slutsky2004BPJ,Lomholt2009PNAS,Zhou2011PNAS,Sheinman2012RPP}, there are also critical reviews focusing on limitations of the traditional model \cite{Mirny2009JPA,Kolomeisky2011PCCP}.

Even if it is accepted by most of the scientists that \emph{in vitro} TFs perform facilitated 
diffusion to find their targets, there is a vivid debate on whether this mechanism indeed plays a 
role \emph{in vivo}.
The interest in this long-standing topic was boosted by the development of new experimental 
techniques, namely single-molecule assays studying DNA-binding proteins, or more generally the
diffusion of proteins within cells
\cite{Sokolov2005BPJ,Gowers2005PNAS,Wang2006PRL,Kolesov2007PNAS,Bonnet2008NAR,vandenBroek2008PNAS,Konopka2006JoB,Kuehn2011PLOS1}.
After finding indirect evidence some years ago, Elf and coworkers recently demonstrated that the lac repressor does display facilitated diffusion in live \emph{Escherichia coli (E. coli)} cells 
\cite{Elf2007Science,Hammar2012Science}.

Thus it is important to study how the present facilitated diffusion models
need to be translated to the \emph{in vivo} situation.  In comparison to
the dilute situation studied \emph{in vitro} the most important changes
are: the influence of the confinement to the cell body or the nucleoid and
the compactified DNA conformation, and the impact of the presence of many
large biomolecules. The latter, which is often referred to as macromolecular
crowding has two major effects: the equilibrium for DNA-binding proteins is
shifted favoring the associated state and the diffusion in the cytoplasm is
slowed down \cite{Minton2001JBC,Morelli2011BPJ}. There is an on-going debate
whether this reduced diffusion is still Brownian, following experimental
evidence that for larger molecules such as mRNA \cite{Golding2006PRL,Weber2010PRL}
or lipid granules \cite{Jeon2011PRL} the motion follows the laws of anomalous
diffusion \cite{Metzler2000PhysRep,Barkai2012PT}. Indeed, there are indications that particles
of the size of several tens of kilo Daltons exhibit anomalous diffusion
\cite{Banks2005BPJ,Weiss2004BPJ}. In what follows we model TFs in the bulk by normal Brownian
diffusion and point at potential implications of anomalous diffusion in the
conclusions.

We note that theoretical work on facilitated diffusion \emph{in vivo}
has also been reported by Mirny and coworkers as well as by Koslover and
coworkers \cite{Mirny2009JPA, Koslover2011BPJ}. A different approach for
the situation in living cells, based on a fractal organization of
the chromatin in the nucleus, showed that also in eukaryotes facilitated
diffusion can be beneficial \cite{Benichou2011PRL}.

With respect to the impact of the cell's finite size Foffano~et~al. recently
studied the influence of (an-)isotropic confinement on the facilitated
diffusion process for rather short DNA chains \cite{Foffano2012PRE}.
To build a theoretical model for facilitated diffusion on the entire genome
in living cells we shortly review what is known about the organization of the
bacterial DNA \cite{Rocha2008ARG}. The emerging general consensus points at a
distinct separation of the genome into connected subunits, that may be dynamic.
 Using atomic force microscopy the size
of structural units of the \emph{E. coli} chromosome was studied, finding units
of size $40$\,nm and $80$\,nm \cite{Kim2004NAR}.  By means of two complementary
approaches the average size of the structural domains was measured to be
$10$ kilobasepairs (kbp) \cite{Postow2004GD}. Romantsov~et~al. studied the structure with
fluorescence correlation spectroscopy, yielding units of size $50$\,kbp with a
diameter of $(70\pm20)$\,nm \cite{Romantsov2007BPJ}.  Chromosome conformation
capture carbon copy(5C) was used to determine a three-dimensional model
of the \emph{Caulobacter} genome \cite{Umbarger2011MolC}.  For the same
bacterium Viollier~et~al. determined that the location of genes on the
chromosome map correlates linearly with its position along the cell's long
axis \cite{Viollier2004PNAS}.

Based on these experimental observations several
models for the DNA structure in bacterial cells have been proposed: entropy
is spotted to be the main driver of chromosome segregation, and ring polymers
are used to model the bacterial chromosome \cite{Jun2010NRMbio,Jung2012SoMa}.
Buenemann and Lenz showed that a geometric model based on a self-avoiding
random walk (SAW) is sufficient to explain the linear positioning of loci
along the cell's longest axis \cite{Buenemann2010PLOS1}.  Finally, the
chromosomal structure and, in particular, the accurate positioning of loci
was proposed as resulting from regulatory interactions \cite{Junier2010PLOSCB,Fritsche2012NAR}.

In this paper we survey if it is possible to extend our previous generalized
facilitated diffusion model \cite{Bauer2012BPJ} to the \emph{in vivo}
situation and compare the results with the ones obtained by Koslover~et~al.
\cite{Koslover2011BPJ}.  Therefore in the following section we detail how
we obtain a coarse-grained model for the bacterial genome and state our
semi-analytical model for the search process.  Then the general theory will
be applied to the specific case of a lac repressor in an \emph{E. coli} cell,
and we favorably compare our results with related experimental measurements
\cite{Elf2007Science}.  Finally we conclude our findings and give an outlook
on future research directions.

\section*{Theory}

The quantity we investigate is the average time a TF needs to find a target sequence in a living
bacterial cell after starting at a random position within the cell.
In principle it is possible to apply our previous generalized diffusion model using rescaled rates,
lengths and diffusion constants to account for the crowded \emph{in vivo} environment \cite{Bauer2012BPJ}.
However, for parameters typical for the interior of cells the effective contact radius between TF 
and DNA is larger than the average distance between neighboring DNA segments.
Consequently a direct translation is not possible.

Moreover, as we will see below, already the simpler one-state model of
facilitated diffusion is sufficient to obtain a fairly good
estimate of the experimental results without any further free parameters.
Thus we do not distinguish between search and recognition states of the TF-DNA
complex \cite{Slutsky2004BPJ}. Intersegmental jumps and/or transfers
\cite{Lomholt2009PNAS,vandenBroek2008PNAS,Sheinman2012RPP,Sheinman2009PhysBiol} of TFs between DNA
segments, that are close-by in the embedding space but distant when measured
in the chemical coordinate along the genome, are to some extent indirectly
included in terms of re-attachment to the DNA within one of the geometric
subunits of the chromosome. In future studies these effects could be explicitly
included to refine the model.

Our approach is based on the general picture of the facilitated
diffusion mechanism:
the TF diffuses three-dimensionally through the bulk solution until it encounters a stretch of DNA to 
which it can bind.
Then a sliding motion along the DNA contour is possible, during which the TF probes for the target.
If the target is not found, the TF will dissociate from the chain after a certain time span and resume
its 3D-diffusion through the cell until the next binding event.
This scheme continues until the target is found.
The major difference to the dilute \emph{in vitro} situation lies in the DNA conformation which is
heavily influenced by the confinement to the cell volume or the nucleoid volume:
As the contour length of (the typically circular) bacterial DNA is about three orders of magnitude 
larger than the longest cell axis in which it is placed, there is clearly a need to compact it.
To proceed we present our model for the compacted genome.

\subsection*{Model for the compacted genome}

Without dwelling on details to which extent nucleoid-structuring proteins and/or supercoiling is 
responsible for DNA compaction in bacterial cells, we adapt the model of Buenemann and Lenz and assume that the DNA is 
assembled structurally into spheres (`blobs') containing one loop each \cite{Buenemann2010PLOS1}.
Thus, the whole genome is modeled as a closed SAW of these uniformly large blobs on a lattice
representing the nucleoid volume (here we diverge from ref. \cite{Buenemann2010PLOS1}, where the 
full cell volume was taken).
To mimic the cylindrical shape of the nucleoid one of the cuboid lattice's edges is taken to be 
longer than the other two of equal length.

\begin{figure}[!h]
\begin{center}
\includegraphics[width=.3\textwidth]{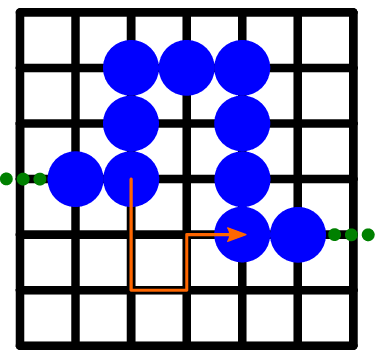}
\end{center}
\caption{{\bf Two-dimensional schematic of the DNA conformation.} The circles denote single DNA
blobs. The lattice spacing is twice the blob radius: $d_g=2r_g$. A part of an 
exemplary search trajectory is depicted by the arrow.
\label{fig:lattice1}}
\end{figure}

The key quantities are the blobs' radius of gyration $r_g$ and the number of basepairs within a
blob, $N_b$.
While the latter parameter determines how many blobs make up the DNA, since the number of bps
on the DNA is a fixed parameter, the first one effectively determines the lattice size
(see figure~\ref{fig:lattice1}).

To obtain individual DNA conformations we follow a routine similar to the one
described in ref.~\cite{Buenemann2010PLOS1}:
as a starting point we use a closed loop of minimal extension which touches both end faces along the
longest cell axis.
Then the chain is elongated by inserting hooks at random positions until it reaches the desired 
length (due to the form of the algorithm only chains with an even number of blobs are considered).
Only elongation steps which yield a conformation within the nucleoid volume are executed.
Afterwards the genome is equilibrated in the following manner: 
we randomly choose one of the three transformation types of the MOS algorithm \cite{Madras1990JSP}. 
Then it is checked if the resulting conformation is still an SAW, otherwise the old conformation is kept. 
Finally only attempts are counted in which the SAW is still confined to the nucleoid volume.
This is repeated 100,000 times for each individual model genome.

Afterwards the resulting DNA conformations are centered on a larger lattice representing the full
cell volume and remain unchanged during the subsequent simulation of the target search process.
This approach is affirmed by recent results that DNA dynamics only have little effect on target
search rates \cite{Koslover2011BPJ}.
For the sake of simplicity we assign the target to be in a blob in the middle
of the DNA.

\subsection*{Target search process}

\begin{figure}[!h]
\begin{center}
\includegraphics[width=.4\textwidth]{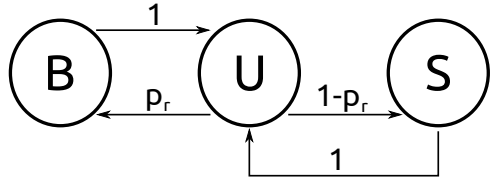}
\end{center}
\caption{{\bf Schematic of the microscopic events within
a blob (without target).} B denotes a bound TF, and U an unbound TF within a blob. Finally, S represents a searching TF which is currently not in a blob.
\label{fig:scheme2}}
\end{figure}

The TF is assumed to start its search at a random position in the cell
volume and its motion is modeled as a random walk on the effective lattice
(fig.~\ref{fig:lattice1}), during
which we keep track of how often sites containing a blob are passed.  The
search process is schematically depicted in fig.~\ref{fig:scheme2}.

The TF starts its search diffusing in 3D (S-state).  With certainty
(probability 1) after some time it will encounter a blob, which it enters
in its unbound state (U).  We first study the case where this blob does
not contain the target DNA.  Based on the microscopic model be
outlined below, we assign a probability $p_r$ that the TF will bind to
the DNA within this blob.  If so it changes to the B-state.  As there is no
target to be found on the DNA, after some time the TF will dissociate and
return to the unbound U-state.  With probability $p_r$ it can bind again, or
it may leave the blob (with probability $1-p_r$) and start a new random walk
on the lattice (S-state).  The same procedure will take place when subsequent
blobs are encountered.

A qualitatively new event occurs when the site containing the target DNA is
encountered for the first time.  Now the tendency to quit the corresponding
blob competes with the probability to find the target. For this reason,
in general
several encounters with the target blob are necessary.  The corresponding
scheme is depicted in figure~\ref{fig:scheme1}:

\begin{figure}[!h]
\begin{center}
\includegraphics[width=.5\textwidth]{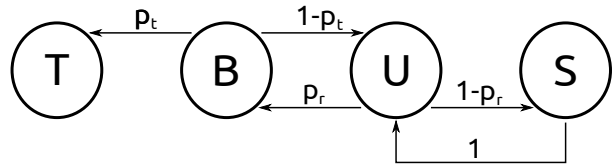}
\end{center}
\caption{{\bf Schematic of the microscopic events within the target blob.} Same notation as in the previous figure. Additionally,
T denotes a TF which has found the target.
\label{fig:scheme1}}
\end{figure}

Once again after entering the blob in the unbound U-state, with probability
$1-p_r$ not a single binding event takes place.  However, if the TF binds to
DNA (with probability $p_r$), subsequently with probability $p_t$
the target will be found (T-state) before dissociating.  If the target
is not found and the TF dissociates, again with probability $1-p_r$, the blob
is left.  Otherwise (with probability $p_r$) a new chance to find the target
while being bound is opened up.  As in the simpler scheme without target,
a new random walk (S-state) is started on a neighboring site if the blob is
left.  To proceed we relate the probabilities $p_r$ and $p_t$ to microscopic
quantities and determine the time steps of the individual processes, before
calculating the typical search time for the target.

\subsection*{Microscopic model}

To determine $p_r$, that is the probability to bind to DNA after entering a blob or after
dissociation from the DNA within the blob we employ the approximation that locally the DNA can
be treated as a random coil \cite{Berg1981BC,Bauer2012BPJ}.
Thus we have to solve the diffusion equation for an initially homogeneous probability distribution
within a sphere of radius $r_g$.
Inside this sphere nonspecific association to a basepair on the DNA occurs with the constant, intrinsic rate $k_{\mathrm{ass}}$ (in units of $M^{-1}s^{-1}$).
We introduce a second concentric sphere of radius $\alpha r_g$ whose
surface is absorbing, modeling the TFs leaving the domain of the blob.
Thus, the dimensionless quantity $\alpha$ measures (in units of $r_g$)
where the blob's area of influence ends, see below and Supporting Information
(SI) S1.
The corresponding problem is solved in the SI S1,
yielding the binding probability
\begin{equation}
 p_r=1-\frac{3\alpha \phi(\gamma)}{\alpha+(\alpha-1)\gamma^2 \phi(\gamma)},
\label{bindprob}
\end{equation}
with the dimensionless quantity $\gamma=r_g\sqrt{\kappa/D_3}$. Here $D_3$
denotes the 3D-diffusion constant, and $\kappa=n k_{\mathrm{ass}} N_b$.
Moreover, $n=3/(4\pi r_g^3)$ represents the density of DNA within the coil. 
In Eq.~\ref{bindprob} we also introduced the auxiliary function
$\phi(\gamma)=(\gamma\coth(\gamma)-1)/\gamma^2$ \cite{Reingruber2010JPCM}.

Note that $p_r$ is a monotonic function of $\gamma$. Keeping
the values of $\kappa, \alpha$ and $r_g $ fixed, for decreasing, yet finite
values of $D_3$ the probability to escape the blob (which is given by $1-p_r$)
becomes smaller, as in this case the TF moves slower and spends more time
within the blob, where it can be caught by a stretch of DNA. Exactly at $D_3=0$
one obtains $p_r=0$, an apparent paradox. However, while it is true that an
immobile TF is unable to leave a blob, the converse argument that the
TF will bind to DNA with certainty is not obvious, as binding requires the
motion of a TF towards DNA within the blob. Because this complementarity is
implicitly assumed in the present model, it only yields meaningful results
for finite values of $\gamma$. Only this situation will be considered in
the following.

If binding occurs, the average time this takes is given by a somewhat
complicated formula for arbitrary values of $\alpha$ (see SI S1).
Here we report the simpler result for the special case $\alpha=2$.
This case is of interest, as in the numerical evaluation
we use the value $\alpha=\sqrt{23/5}\approx2.14$, see below.
\begin{equation}
 \tau_b^{\alpha=2}=\frac{1}{2\kappa}\frac{20+(8\gamma^2-30)\phi(\gamma)+(4\gamma^2-36)
\gamma^2\phi^2(\gamma)}{(2+\gamma^2\phi(\gamma))(2+(\gamma^2-6)\phi(\gamma))}.
\end{equation} 
Conversely, the average time it takes the TF to leave the blob is
\begin{equation}
 \tau_e^{\alpha=2}=\frac{1}{2\kappa}\frac{6-2\phi^{-1}(\gamma)+\gamma^2(4\phi(\gamma)+\frac{4}{3})+
\frac{\gamma^4}{3}\phi(\gamma)}{2+\gamma^2\phi(\gamma)}.
\end{equation}
While diffusing in 3D, a single random walk step on average takes $\tau_\mathrm{3D}=d_g^2/(6D_3)$.
Once the TF binds non-specifically to the blob containing the target, the probability to find the
target before dissociating can be found by considering a one-dimensional diffusion problem. 
We assume that the target is located in the middle of the corresponding blob.
Then we consider a DNA stretch of length $L=N_bb/2$ with the target at one end.
Here $b$ denotes the size of a basepair, $b=0.34$ nm.

Due to the DNA's coiled conformation within a blob, we use the standard assumption
that the first binding event occurs at a random position on the DNA and that dissociation and
reassociation positions are completely uncorrelated, see for example \cite{Coppey2004BPJ}.
Formally this implies that the TF initially is uniformly distributed on the DNA along which it diffuses
with the diffusion constant $D_1$.
The TF can leave the DNA with the dissociation rate $k_\mathrm{off}$.
We furthermore assume that the other extremity of the DNA acts as a reflecting boundary\cite{Coppey2004BPJ}, possibly due
to compacting proteins that obstruct further 1D-diffusion at this position.
The calculation detailed in the SI S1 yields:
\begin{equation}
 p_t=\frac{\tanh(L/\ell)}{L/\ell},
\label{pt}
\end{equation}
with $\ell=\sqrt{D_1/k_\mathrm{off}}$, which denotes a typical distance covered sliding on DNA before dissociating. If the target is found, the conditional time this successful event takes on average, reads
\begin{equation}
 \tau_t=\frac{1-1/\left(p_t\cosh^2\left(\frac{L}{\ell}\right)\right)}{2k_\mathrm{off}}=
\frac{1-\frac{L}{\ell}/\left(\sinh\left(\frac{L}{\ell}\right)
\cosh\left(\frac{L}{\ell}\right)\right)}{2k_\mathrm{off}}.
\label{taut}
\end{equation}
However, an unsuccessful event implies that the DNA is (on average) left after the time span
$\tau_d=1/k_\mathrm{off}$.
Inspection of Eq. (\ref{taut}) shows that in the limit
$D_1\rightarrow 0$, i.e. when TFs are (nearly) incapable of sliding,
$\tau_t$ approaches the finite value $1/(2k_\mathrm{off})$, which is at
first sight a surprising result. However, in this limit the probability to
reach the target as given by Eq. (\ref{pt}) approaches zero, ensuring that
meaningful results are obtained. It should be stressed that our model only
allows target detection via sliding, and not via direct detection solely
through three-dimensional diffusion.

\subsection*{Mean search time}

To determine the mean time it takes to find the target at first we specify how often
the ``loop'' of binding and unbinding events (B and U in figures~\ref{fig:scheme2} and
\ref{fig:scheme1}) is executed during an encounter with a blob.
In all the blobs without the target this happens on average $p_r/(1-p_r)$ times.
As one loop lasts $\tau_c=\tau_b+\tau_d$ the average time that is spent within a
blob is $\tau_\mathrm{blob}=\tau_e+\tau_c p_r/(1-p_r)$.

In the blob containing the target, the average number of binding and unbinding loops is
$g(p_r,p_t)=\chi/(1-\chi)$, where $\chi=p_r(1-p_t)$.
Note that the number of executed loops in blobs without target is the special case $p_t=0$ of the
general case, $g(p_r,p_t=0)=p_r/(1-p_r)$.
In the same sense figure~\ref{fig:scheme2} can be considered a special case of
figure~\ref{fig:scheme1}.
The combined probability to find the target before leaving the blob reads
$p_rp_t/(1-p_r+p_rp_t)$, consequently the probability for a failed attempt is
$p_\mathrm{uns}=(1-p_r)/(1-\chi)$.
Thus, a successful event during which the target is found, on average takes
$\tau_\mathrm{suc}=\tau_b+\tau_t+g(p_r,p_t)\tau_c$, and an unsuccessful one
$\tau_\mathrm{uns}=\tau_e+g(p_r,p_t)\tau_c$.

The mean total search time can be dissected into three contributions:
first, the mean time the TF needs to arrive at the target blob for the first time.
Then the mean time it takes to return to the target after an unsuccessful search event.
The latter has to be multiplied with the average number of failed attempts.
Finally the average time it takes to successfully bind the target at the
corresponding blob has to be added.

To quantify this model two parameter pairs from the random walk simulation are
needed as inputs:
the mean number of steps it takes to encounter the target blob for the first time
$n_\mathrm{f,3D}$ after starting at a random position within the cell and how many
blobs without target are encountered during this time, $n_\mathrm{f,enc}$.
Furthermore we determine the mean number of steps and blob-encounters in a random walk
starting on a site next to the target blob: $n_\mathrm{r,3D},n_\mathrm{r,enc}$ and
ending in the target blob. Altogether the mean total search time reads:
\begin{eqnarray}
\nonumber
\tau&=&n_\mathrm{f,3D}\tau_\mathrm{3D}+n_\mathrm{f,enc}\tau_\mathrm{blob}\\
\nonumber
&&+\frac{p_\mathrm{uns}}{1-p_\mathrm{uns}}\left(\tau_\mathrm{uns}+n_\mathrm{r,3D}\tau_\mathrm{3D}
+n_\mathrm{r,enc}\tau_\mathrm{blob}\right)\\
&&+\tau_\mathrm{suc}.
\label{totaltime}
\end{eqnarray}
This formula is the main result of our study, which will be discussed
quantitatively for the case of the lac repressor in an \emph{E. coli} cell.

\section*{Results}

As input parameters for our TF search model in a living cell we use data deduced from experimental
studies.
For the DNA configuration we use two parameter sets for the blob size and the number $N_b$ of
basepairs within a blob:
(a) $r_g=15$\,nm and $N_b=10^4$  \cite{Postow2004GD,Buenemann2010PLOS1} and (b) $r_g=35$\,nm and
$N_b=5\times10^4$ \cite{Romantsov2007BPJ}. 
The volume of the nucleoid can be approximated as a cylinder of diameter 
$d_{nuc}=0.24\,\mu m$ and length $l_{nuc}=1.39\,\mu m$ \cite{Jun2010NRMbio}.
We use a cuboid with edge lengths $l_x=l_y=\sqrt{\pi\times d_{nuc}^2/4}\approx213\,nm$ and
$l_z=l_{nuc}$.
This corresponds to nucleoid lattices of size $7\times7\times46$ and $3\times3\times20$.
As the \emph{E.coli} genome consists of $\sim4639$\,kbps, we compose a closed
SAW consisting of (a) 464 blobs and (b) 92 blobs, respectively.
For the parameter sets we create three and five sample conformations.
The total cell volume can be approximated as a cylinder with $d_{cell}=0.5\,\mu m$ and length 
$l_{nuc}=2.5\,\mu m$ \cite{Jun2010NRMbio}.
Accordingly, we use embracing lattices of size $15\times15\times83$ and $6\times6\times36$ to
mimic the full cell volume.
Besides, we employ $\alpha=\sqrt{23/5}$ in order to obtain the correct asymptotic behavior
for small values of $k_\mathrm{ass}$ as detailed in the SI S1 and we use $D_3=3\mu m^2/s$ and
$D_1=0.046\mu m^2/s$ \cite{Elf2007Science}.
The results of the random walk simulation are summarized in table~\ref{simresults}.

\begin{table}[!ht]
\caption{\bf{Simulation results}}
\begin{center}
\begin{tabular}{|c|c|c|c|c|c|c|c|c|}
\hline
Set & $n_\mathrm{f,3D}$ & $n_\mathrm{f,enc}$ &$q_f$& $n_\mathrm{r,3D}$ &
$n_\mathrm{r,enc}$& $q_r$\\
\hline
a & $31514$ & $766.41$ & $0.0243$& $18689$ & $463.48$& $0.0248$\\
\hline
b & $2594.7$ & $175.63$ & $0.0677$&$1291.9$ & $90.848$& $0.0703$\\
\hline
\end{tabular}
\end{center}
\begin{flushleft}
\begin{center}
Simulation results for parameter sets a and b
\end{center}
\end{flushleft}
\label{simresults}
\end{table}

A first inspection of the values of $n_\mathrm{r/f,3D}$ and $n_\mathrm{r/f,enc}$
shows that the ones obtained with parameter set a are approximately one order of
magnitude larger than the ones obtained with set b. This is clear as set a
corresponds to a finer model of the DNA, in which the respective value of $r_g$
is smaller. Next, we consider the ratios $q_f=n_{f,enc}/n_{f,3D}$ and $q_r=n_{r,
enc}/n_{r,3D}$, that is the fractions of sites containing a blob encountered
during a trajectory. The results are very close to the total fraction of sites
that are occupied by a blob: for parameter set a, this is: $464/(15\times15
\times83)\approx0.0248$ and for b: $92/(6\times6\times36)\approx0.0710$.
This and the fact that the values for the first encounter and for the returning
trajectories are similar, support the statement that the TF experiences an
effective medium through which it diffuses \cite{Koslover2011BPJ}. If we only
consider the mean search times, this medium is mainly characterized by the
mean DNA density within the cell.

\subsection*{Non-monotonic behavior}

\begin{figure}[!h]
\begin{center}
\includegraphics[width=4in]{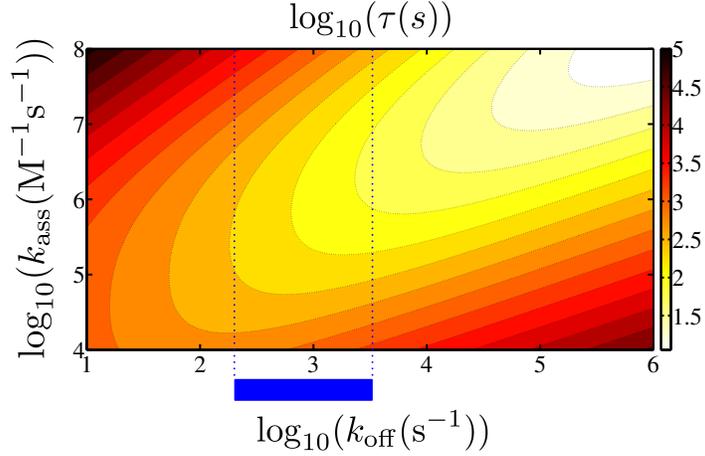}
\end{center}
\caption{{\bf Mean search time.}  The mean search time is plotted as a function of the dissociation rate $k_\mathrm{off}$ and
the association rate $k_\mathrm{ass}$ (using parameter set a). The blue
bar and the blue dotted lines denote the range of $k_\mathrm{off}$ which is biologically
relevant \cite{Elf2007Science}.
\label{fig:plot1}}
\end{figure}

In figure~\ref{fig:plot1} the mean search time averaged over the
ensembles with parameter set a is shown as a function of the association rate
$k_\mathrm{ass}$ and the dissociation rate $k_\mathrm{off}$.

We find a non-monotonic dependence both on the association and the dissociation rate
typical for facilitated diffusion models:
for a fixed value of $k_\mathrm{ass}$ there exists a value of $k_\mathrm{off}$ that minimizes 
the search time.
This minimal value decreases if both rates are increased while keeping them at a constant
ratio.

\begin{figure}[!h]
\begin{center}
\includegraphics[width=4in]{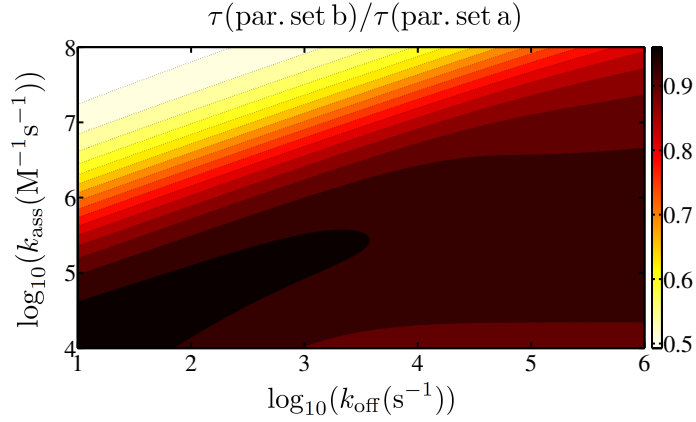}
\end{center}
\caption{{\bf Difference between the two parameter sets.} The plot shows the ratio of the mean search time obtained with parameter set b with the
ones obtained with set a.
\label{fig:plot2}}
\end{figure}

In figure~\ref{fig:plot2} the ratio of the search time obtained with parameter set b
with the search time obtained with parameter set a is plotted for the same range
as in figure~\ref{fig:plot1}.

Even though set b always yields slightly smaller search times, the results are very similar,
especially in the range usually studied in experiments, as we will see below.
Therefore in the following we solely consider results obtained with set a.
In the SI S1 we moreover show that the approach to use an ensemble average to obtain the mean
search time is justified as the scatter between data obtained with individual conformations
is negligible (see figure S1). Only at very low values of $k_\mathrm{off}$, when the TF spends
considerable time in the non-specifically bound state,
the individual conformation does play a role.

We saw that for fixed values of $k_\mathrm{ass}$, there exists an optimal
of $k_\mathrm{off}$, for which the target localization occurs fastest.
It is insightful to study whether a living \emph{E. coli}
cell operates close to this point.

\subsection*{Comparison to experimental results}

\begin{figure}[!h]
\begin{center}
\includegraphics[width=4in]{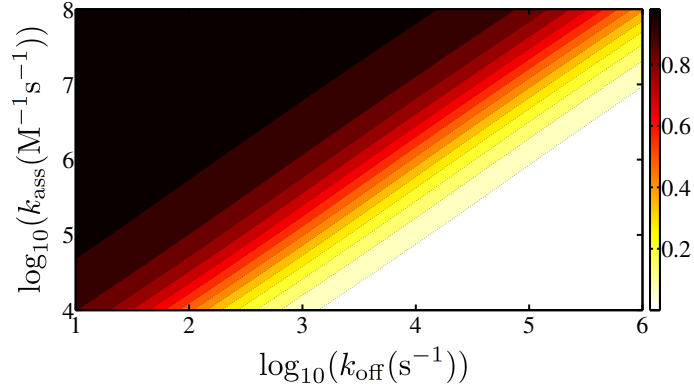}
\end{center}
\caption{{\bf Bound fraction of time.} The fraction of time during which the TF is non-specifically bound is shown (using parameter set a).
\label{fig:plot4}}
\end{figure}

We choose the rates according to the results of Xie and coworkers \cite{Elf2007Science}:
they measured that the lac repressor spends 87\% of the total time non-specifically bound and
determined the residence time on DNA $t_R$ to be in the range
\begin{equation}
 0.3\mathrm{ms}<t_R=1/k_\mathrm{off}<5\mathrm{ms}.
\label{expkoff}
\end{equation}
To incorporate these values, we calculate the fraction of time, $f_b$, that the TF spends
non-specifically bound. This is obtained from Eq.~\ref{totaltime} by only considering the terms involving
$\tau_d$ and $\tau_t$.
The result is plotted in figure~\ref{fig:plot4}, again as a function of dissociation and
association rate.

\begin{figure}[!h]
\begin{center}
\includegraphics[width=3in]{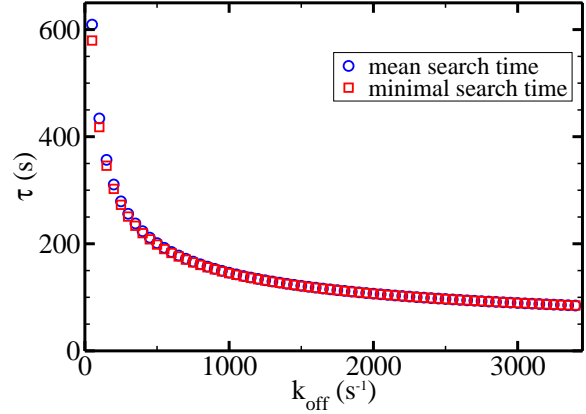}
\end{center}
\caption{{\bf Mean search time and minimal search time.} The mean search time and the minimal search time (with appropriately chosen $k_\mathrm{ass}$)
are plotted as a function of the dissociation rate at parameters relevant for the interior of living cells.
\label{fig:plot7}}
\end{figure}

We see that contour lines of a constant fraction appear as straight lines in this log-log-plot.
A numerical analysis yields that the condition $f_b=0.87$ is fulfilled for
\begin{equation}
 \log_{10}(k_\mathrm{ass}(\mathrm{M}^{-1}\mathrm{s}^{-1}))=1.04
\log_{10}(k_\mathrm{off}(\mathrm{s}^{-1})) + 2.76.
\label{fracbound}
\end{equation}
The observation that the slope of this curve is (nearly) unity, reflects the fact that specifying
the bound fraction of time is equivalent to specifying the equilibrium binding constant which is
simply given by the ratio of $k_\mathrm{ass}$ and $k_\mathrm{off}$.
We plug Eq.~\ref{fracbound} into our model and plot the resulting mean search time
as a function of the single residual parameter $k_\mathrm{off}$ in figure~\ref{fig:plot7} in the
range given by Eq.~\ref{expkoff}. Additionally, in figure~\ref{fig:plot7} we plot the minimal search
time in this regime which is obtained by choosing the optimal value of $k_\mathrm{ass}$.

In both cases we obtain a monotonically decreasing function of $k_\mathrm{off}$.
Most interestingly, the values obtained in this biologically relevant parameter regime are
only marginally larger than the optimal ones.
At $k_\mathrm{off}\gtrsim500\mathrm{s}^{-1}$ the two data sets nearly lie on top of each other.
This means that within our model an \emph{E. coli} cell seems to operate quite close to conditions,
which are optimal for target localization.
At $k_\mathrm{off}=200\mathrm{s}^{-1}$, which was used in the discussion in 
ref.~\cite{Koslover2011BPJ}, we obtain $\tau\approx311$ s.
This is approximately $12$\% below the experimental result $6\times59\mathrm{s}
=354\mathrm{s}$ \cite{Elf2007Science}, implying a very favorable agreement.

\subsection*{Local searches}

\begin{figure}[!t]
\begin{center}
\includegraphics[width=4in]{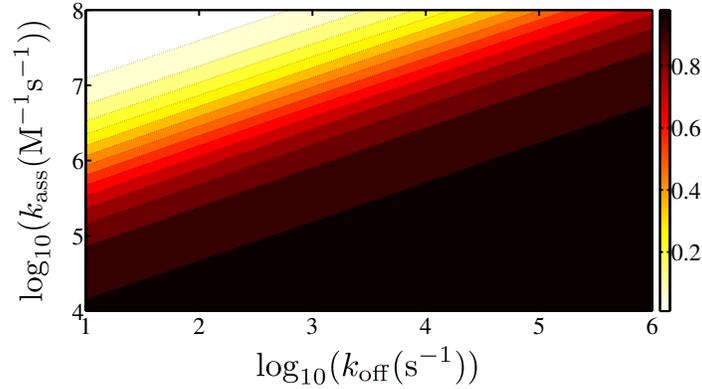}
\end{center}
\caption{{\bf Acceleration due to local searches.} The ratio of the time needed in a local search with the one in a global search (with 
parameter set a) is shown.
\label{fig:plot6}}
\end{figure}

There is some evidence that many TFs are produced close to their target positions, a phenomenon
called colocalization \cite{Kolesov2007PNAS,Wunderlich2008NAR}.
These local searches would obviously be faster than a global search starting at a random position
within the cell.
To quantify this in figure~\ref{fig:plot6} we plot how many percent of the total search time is
still needed to find the target if the TF starts its search in the target blob while all other
parameters remain unchanged.

In mathematical terms this corresponds to omitting the terms in the first line
of Eq.~\ref{totaltime}.
We see that only for relatively large values of $k_\mathrm{ass}$ an appreciable acceleration is
obtained for local searches.
This is clear as large values of the association rate imply that all the blobs encountered en
route act as traps slowing down the transport.
Interestingly, in the regime typical for the interior of cells the acceleration is of little amount.
This can also be interpreted in the more general context of ``geometry-controlled
kinetics'', see the works of B\'{e}nichou and coworkers \cite{Benichou2010NatChem,Meyer2011PRE}.
These authors showed that for non-compact exploration of space - as is the case in the present model -
the initial position of a searching particle has little influence.

\section*{Discussion}

We analyzed the facilitated diffusion mechanism in a living cell using a coarse-grained
model of the bacterial genome.
Just like in dilute \emph{in vitro} systems there is a non-monotonic dependence both on
the dissociation rate and the association rate of TFs from and to DNA.
The respective optimal conditions mark a trade-off between spending too much time on DNA where
the motion is rather slow, but the target can be found, and spending too much time
in the cytoplasm where the motion is faster, but the TF is insensitive to the target.

When calculating the mean search time as an input from our random walk simulation we solely
use the mean number of steps taken and the number of blobs encountered during the trajectory.
This corresponds to treating the nucleoid body as an effective medium through which the
TF diffuses, which agrees with the observations made by Koslover~et~al. that within a short
time span the TF starts an effective diffusive motion \cite{Koslover2011BPJ}.
Accordingly, we see that the exact values of the parameters describing the DNA conformation
have only little effect on our results.
Only the fact that there is an effective medium
characterized by the DNA density matters.

Calibrating our results with the experimental observation that the TF spends
87\% of the time
non-specifically bound \cite{Elf2007Science}, we obtain search times that only
slightly underestimate the experimentally known results.
In a previous study we showed that the introduction of a search and a recognition state
in order to resolve the speed-stability paradox slows down the search \cite{Bauer2012BPJ}.
Thus, a refined model taking this effect into account could yield a result even
closer to the experimental one.

Most importantly, within our model the results in the biologically relevant regime of 
dissociation rates are quite close to the ones minimizing the search time, indicating that
living \emph{E. coli} cells function near conditions optimal for TF target location.

Our results for the mean search times are similar to those obtained by
Koslover et al.~\cite{Koslover2011BPJ}. However, in their model for
\emph{in vivo\/} facilitated diffusion they distribute the DNA over the
entire cell volume and assume a random coil configuration.
If one were confining the DNA to
the smaller nucleoid volume, the effective DNA-TF contact radius in that
model would then become smaller than the average distance between DNA segments.
Besides, our model is less idealized.
In that sense our current approach has the advantage that the DNA is
realistically confined to the nucleoid volume, and based on input
parameters deduced from experimental studies we also obtain mean search times, that are very close to experimental
\emph{in vivo\/} values. Moreover, our model offers the advantage that
in future studies additional information may be deduced, for example,
by studying the underlying probability densities of $n_\mathrm{r,3D}$,
$n_\mathrm{r,enc}$, etc., in addition to their mean values determined here.

\subsection*{Colocalization effects}

Comparing the mean search times for TFs starting at a random position in the cell volume with
those TFs that already start close to the target, we only observe a minor acceleration.
This is due to the fact that most of the search time is spent returning to the target blob after a failed
attempt to find the target.
For a wide range of parameters the first encounter with the target blob only represents
a small fraction of the whole search time.
Leaving the picture of mean values for the search time of an ensemble of TFs,
on the level of single trajectories immediate returns to the target blob are
indeed possible and thus may lead to search times much shorter than the average
search time. Such scenarios may in fact be relevant for biological cells.

Should observations of anomalous diffusion for TFs in the cytoplasm of living
cells be substantiated, the effect of colocalization should become 
significantly more pronounced, if the nature of the exploration of space is compact
\cite{Benichou2010NatChem,Meyer2011PRE}: subdiffusion implies an increased occupation
probability near the initial position
\cite{Golding2006PRL,Guigas2008BPJ,Lomholt2007PRL}, and
thus increases the likelihood for successful TF-DNA binding after repeated
attempts. In that sense subdiffusion may even be beneficial for molecular
processes in living cells, as argued recently
\cite{Guigas2008BPJ,weiss,leila}.\\

We believe that this relatively simple model for facilitated diffusion in
vivo will instigate new experiments and more detailed theories, to
ultimately obtain a full understanding of bacterial gene regulation.

\section*{Acknowledgments}
This work was supported by Academy of Finland (FiDiPro scheme): www.aka.fi/eng; and German Federal Ministry for Education and Research: www.
bmbf.de/en/index.php.

\clearpage

\section*{Supporting Information S1}
\renewcommand{\thefigure}{S\arabic{figure}}
\renewcommand{\theequation}{S\arabic{equation}}
\setcounter{equation}{0}
\setcounter{figure}{0}
In this supporting information we detail the explicit calculations which are beyond the scope
of the main text.

\section{Microscopic model}
\label{sec:micrmod}

\subsection{Association probability}
\label{sec:assprob}
To relate $p_r$ to the non-specific association rate $k_{\mathrm{ass}}$ per base pair
(in units of $M^{-1}s^{-1}$), we solve the following diffusion equation for the TF's
probability $c(\textbf{r},t)$ to be at position \textbf{r} at time t:
\begin{equation}
 \frac{\partial c(\textbf{r},t)}{\partial t}= \left\{\begin{array}{cl}
D_3\Delta c(\textbf{r},t)-\kappa c(\textbf{r},t), & \mbox{for }0<r<r_g\\
D_3\Delta c(\textbf{r},t), & \mbox{for }r_g<r<r_2 \end{array}\right.,
\label{diffeq1}
\end{equation}
with $\kappa=n k_{\mathrm{ass}} N_b$, where $n$ denotes the density of DNA and $N_b$ the
number of basepairs within the blob.
$D_3$ denotes the 3D-diffusion constant and $r_g$ the blob's radius of gyration.  
The differential equation is subject to the initial condition 
\begin{equation}
 c(\textbf{r},t=0)=\left\{\begin{array}{cl} c_0=3/(4\pi r_g^3), & \mbox{for }0<r<r_g\\ 0, &
\mbox{for }r_g<r<r_2 \end{array}\right. ,
\end{equation}
and the boundary condition $c(r=r_2,t)=0$.
Thus, $r_2$ represents a cutoff-radius at which the TF is assumed to have
definitely left the domain of the blob.
We use $n=c_0$ as we study the situation where one TF is in the blob containing one
DNA chain.

We define the Laplace transform $f(u)$ of a function $f(t)$ through:
\begin{equation}
f(u)=\int\limits_0^{\infty}f(t)\exp(-ut)dt.
\end{equation}

In Laplace space the differential equation \ref{diffeq1} reads:
\begin{equation}
 u c(u,r)=\left\{\begin{array}{cl} c_0+D_3\Delta c(\textbf{r},u)-
\kappa c(\textbf{r},u), & \mbox{for }0<r<r_g\\
D_3\Delta c(\textbf{r},u), & \mbox{for }r_g<r<r_2 \end{array}\right. ,
\end{equation}
From its solution the flux out of the outer sphere $j_{\mathrm{out}}(u)$ and the
binding
flux $j_\mathrm{bind}(u)$ in the inner sphere can be obtained via:
\begin{equation}
 j_{\mathrm{out}}(u)=-4\pi r_2^2D_3\left.\frac{\partial
c(u,r)}{\partial r}
\right|_{r=r_2},
\end{equation}
and
\begin{equation}
 j_\mathrm{bind}(u)=4\pi\kappa\int\limits_0^{r_g}dr\, r^2 c(u,r).
\end{equation}

We obtain
\begin{equation}
 j_{\mathrm{out}}(u)=\frac{3}{r_g^3q_1^3}\frac{r_2q_2}{\sinh(q_2\delta r)}
\frac{q_1r_g\coth(q_1r_g)-1}{\coth(q_1r_g)+\frac{q_2}{q_1}\coth(q_2\delta r)},
\end{equation}
and furthermore
\begin{align}
 j_\mathrm{bind}(u)=\frac{3}{r_g^3q_1^3}\frac{\kappa}{u+\kappa}\left[\frac{r_g^3q_1^3}{3}\right.\notag\\
\left.-\frac{(q_1r_g\coth(q_1r_g)-1)(1+r_gq_2\coth(q_2\delta r))}{\coth(q_1r_g)+\frac{q_2}{q_1}\coth(q_2\delta r)}\right],
\end{align}
where $q_1=\sqrt{\frac{u+\kappa}{D_3}}$, $q_2=\sqrt{u/D_3}$ and $\delta r=r_2-r_g$.

A Taylor series around $u=0$ then yields
\begin{equation}
j_\mathrm{bind}(u)\simeq p_r(1-\tau_b u),
\end{equation}
and
\begin{equation}
j_\mathrm{out}(u)\simeq(1-p_r)(1-\tau_e u).
\end{equation}

We obtain
\begin{equation}
 p_r=1-\frac{3\alpha \phi(\gamma)}{\alpha+(\alpha-1)\gamma^2 \phi(\gamma)},
\end{equation}
where we introduced $\alpha=r_2/r_g$, $\gamma=r_g\sqrt{\kappa/D_3}$ and
the auxiliary function
$\phi(\gamma)=(\gamma\coth(\gamma)-1)/\gamma^2$ [S1].

The average time it takes for binding reads
\begin{align}
 \tau_b=\frac{\alpha}{2\kappa}\left\{5\alpha+\left(4\gamma^2(\alpha-1)-15\alpha\right)\phi(\gamma)\right.\notag\\
\left.+\left(12-15\alpha+2\gamma^2(1-\alpha)^2\right)\gamma^2\phi^2(\gamma)\right\}\notag\\
\times\left(\alpha+(\alpha-1)\gamma^2\phi(\gamma)\right)^{-1}\notag\\
\times\left(\alpha+(\gamma^2(\alpha-1)-3\alpha)\phi(\gamma)\right)^{-1}.
\end{align}
This equation is true for arbitrary values of $\alpha$.
In the main text we explicitly state the case $\alpha=2$. However, in the results section
we use $\alpha=\sqrt{23/5}\approx2.14$, as described in the last section of this SI.

The average time the TF needs for leaving the blob is given by
\begin{eqnarray}
 \tau_e=\frac{1}{2\kappa}\left\{\alpha(3-\phi^{-1}(\gamma))+\gamma^2\left((3\alpha-2)\phi(\gamma)\right.\right.\notag\\
\left.\left.+\frac{2+\alpha}{3}(1-\alpha)^2\right)-\frac{\gamma^4}{3}(1-\alpha)^3\phi(\gamma)\right\}\notag\\
\times(\alpha+(\alpha-1)\gamma^2\phi(\gamma))^{-1}.
\label{tauesc}
\end{eqnarray}

\subsection{Target finding probability}
\label{sec:targprob}
To calculate the probability to find the target before dissociating, we consider the
one-dimensional diffusion problem
\begin{equation}
 \frac{\partial c(z,t)}{\partial t}=D_1\frac{\partial^2c(z,t)}{\partial z^2}-k_{\mathrm{off}}c(z,t),
\end{equation}
subject to the initial condition $c(z,t=0)=1/L$ and the boundary conditions
$c(z=0,t)=0$ and $\left.\frac{\partial c(z,t)}{\partial z}\right|_{z=L}=0$.
In Laplace space with respect to time we obtain the following solution:
\begin{equation}
 c(u,z)=\frac{1}{L(u+k_{\mathrm{off}})}\left(1-\frac{\cosh((L-z)
\sqrt{\frac{u+k_{\mathrm{off}}}{D_1}})}{\cosh(L\sqrt{\frac{u+k_{\mathrm{off}}}{D_1}})}\right)
\end{equation}
A Taylor series of $j_\mathrm{target}(u)=D_1\left.\frac{\partial
c(z,u)}{\partial z}\right|_{z=0}$ in $u$ yields:
\begin{equation}
 j_\mathrm{target}(u)\simeq\frac{\tanh(L/\ell)}{L/\ell}+\frac{u}{2k_\mathrm{off}}
\left(\frac{1}{\cosh^2(L/\ell)}-\frac{\tanh(L/\ell)}{L/\ell}\right),
\end{equation}
where $\ell=\sqrt{D_1/k_{\mathrm{off}}}$.

This corresponds to a target finding probability of
\begin{equation}
 p_t=\frac{\tanh(L/\ell)}{L/\ell}.
\end{equation}
The average time it takes to find the target reads
\begin{equation}
 \tau_{t}=\frac{1}{2k_\mathrm{off}}\left(1-\frac{L/\ell}{\sinh(L/\ell)\cosh(L/\ell)} \right).	
\end{equation}

\section{Justification for the use of the ensemble average}
\label{sec:ensavrg}
\begin{figure}
\begin{center}
\includegraphics[width=4in]{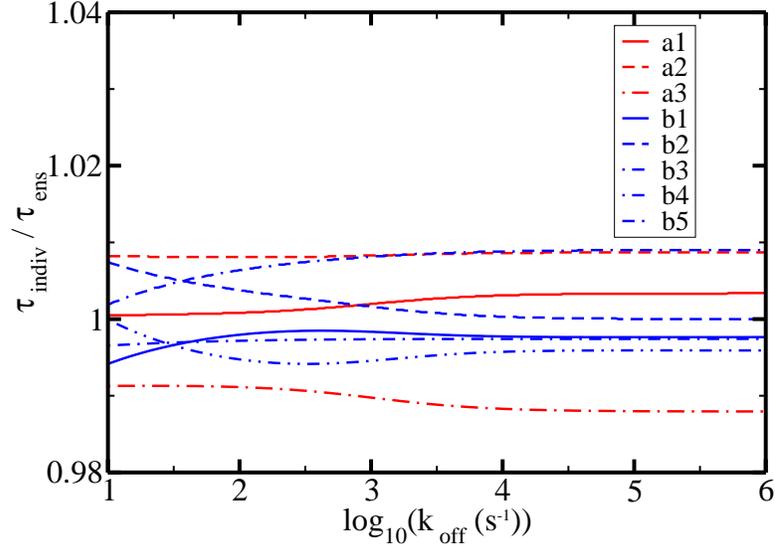}
\caption{Ratio of the mean search times obtained with individual conformations with the respective 
ensemble averaged mean search time at $k_\mathrm{ass}=10^5\mathrm{M}^{-1}\mathrm{s}^{-1}$.}
\label{fig:plot3}
\end{center}
\end{figure}

In Figure \ref{fig:plot3} we plot the ratio of the mean search time for all the eight individual
conformations with the mean search time of the corresponding ensemble average at 
$k_\mathrm{ass}=10^5\mathrm{M}^{-1}\mathrm{s}^{-1}$.

Apparently all the individual curves only scatter about one percent around the value obtained
with the ensemble average. Thus it appears appropriate always to use the latter in the main text.

\section{Derivation of $\alpha=\sqrt{23/5}$}
\label{sec:sqrt235}
In principle the parameter $\alpha$ which represents the ratio of the cutoff-radius $r_2$ and
the blob's radius of gyration $r_1$ is a free parameter which can be used to refine the model.
However, in the limit $\kappa\rightarrow0$, that is when no binding to DNA occurs or when there
is no DNA present, the escape time $\tau_e$ from a blob should coincide with the free diffusion
time $\tau_{3D}$.
Now using Eq. \ref{tauesc},
\begin{equation}
 \lim\limits_{\kappa\rightarrow0}\tau_e(\kappa)=\frac{r_g^2}{30D_3}(5\alpha^2-3).
\end{equation}
Equalizing this with $\tau_{3D}=\frac{4r_g^2}{6D_3}$ yields $\alpha=\sqrt{\frac{23}{5}}$. Consequently,
this value was chosen in the main text.
 
%\bibliographystyle{unsrt}
%\bibliography{/home/t30/met/ga73vin/Latex-Files/fd1}

\section*{References}

S1. Reingruber J, Holcman D
(2010) Narrow escape for a stochastically gated brownian ligand.
{\it J Phys Condens Matter} 22:065103.

\end{document}